\pdfoutput=1

\documentclass[aps,pra,10pt,twocolumn,showpacs,superscriptaddress]{revtex4-1}
\usepackage{amsmath}
\usepackage{latexsym}
\usepackage{pgfplots}
\usepackage{rotating}
\usepackage{amssymb}
\usepackage{bm}
\usepackage{graphics,epstopdf}
\usepackage{color}\usepackage{amsmath}

\usepackage[usenames,dvipsnames,svgnames]{pstricks}
\usepackage{epsfig}
\usepackage{pst-grad} % For gradients
\usepackage{pst-plot} % For axes

\usepackage{newlfont}
\usepackage{amssymb}
\usepackage{amsfonts}
\usepackage{amsmath}
\usepackage{mathrsfs}
\usepackage{amsthm}
\usepackage{graphicx}
\usepackage{epsfig}
\usepackage{color}
\usepackage{bm}
\usepackage[breaklinks]{hyperref}
\usepackage{tikz}

\usepackage{times}

\begin{document}

\title{Generalized Fubini-Study Metric and Fisher Information Metric}
\author{Debasis Mondal}
\email{debamondal@hri.res.in}

\affiliation{Quantum Information and Computation Group,\\
Harish-Chandra Research Institute, Chhatnag Road, Jhunsi, Allahabad, India}


\date{\today}

\begin{abstract}

  We provide an experimentally measurable local gauge $U(1)$ invariant Fubini-Study (FS) metric for mixed states. Like the FS metric for pure states, it also captures only the quantum part of the uncertainty in the evolution Hamiltonian. We show that this satisfies the quantum Cramer-Rao bound and thus arrive at a more general and measurable bound. Upon imposing the monotonicity condition, it reduces to the square-root derivative quantum Fisher Information. We show that on the Fisher information metric space dynamical phase is zero. A relation between square root derivative and logarithmic derivative is formulated such that both give the same Fisher information. We generalize the Fubini-Study metric for mixed states further and arrive at a set of Fubini-Study metric---called $\alpha$ metric. This newly defined $\alpha$ metric also satisfies the Cramer-Rao bound. Again by imposing the monotonicity condition on this metric, we derive the monotone $\alpha$ metric. It reduces to the Fisher information metric for $\alpha=1$.

\end{abstract}

\maketitle

%%%%%%%%%%%%%%%%%%%%%%%%%%%%%%%%%%%%%%%%%%%%%%%%%%%%%%%%%%%%%%%%%%%%%%%%%%%%%%

\section{Introduction}
%%%%%%%%%%%%%%%%%%%%%%%%%%%%%%%%%%%%%%%%%%%%%%%%%%%%%%%%%%%%%%%%%%%%%%%%%%%%%%
With the advent of quantum information theory and the precision measurement techniques various experiments are being set up to explore the mysteries of nature. There are theories like general relativity and quantum mechanics which fit well with all the experimental results and predict a number of physical phenomena in their respective regimes. But the two theories do not fit well with each-other in the interfacing regime where the energy and length scales are of the Planck order.  Recently developed quantum information theoretic tools, particularly, metrology in curved space time scenario have been used to explore and understand phenomena in this regime. A number of experiments requiring the precision measurement techniques have been proposed to understand the nature in this length and energy scale. Quantum Fisher information is the one among others which has widely and successfully been used in quantum information theory, precision measurement and metrology. This is due to the fact that they are monotonic and 
satisfy the Cramer-Rao bound \cite{amari,petz,fisher,cramer,amari1,fuji,matsumoto}.

It is well understood and believed that any successful theory must be a local gauge theory. Twentieth century has seen a rapid advancement in particle physics which is nothing but a systematic progress in unification of the fundamental forces of nature via local gauge theory. It has also seen a number of forecasts and predictions on fundamental particles. Keeping this in mind, we here put forward  a new metric for mixed state which is local gauge invariant. 

The motivation behind such metric is very simple. Suppose a system is evolving. We believe that even if the system is not in a pure state, its purified version \cite{nielsen,karol} in the enlarged Hilbert space, i.e., the system and the environment together evolves satisfying an equation of motion which is local gauge invariant. Therefore, the distance must always be local gauge invariant. There is a unique local gauge invariant metric on the pure quantum state known as the Fubini-Study(FS) metric \cite{pati,anandan,karol}. We call this newly derived metric as FS metric for mixed states which also satisfies the Cramer-Rao bound \cite{petz,amari,amari1,cramer,fisher,fuji,matsumoto}.

Due to the non-commutativity of the quantum states, uniqueness of the classical Fisher information metric does not hold in case of quantum Fisher information metric. We can define infinite number of quantum Fisher information metric on the quantum state space. But none of these metrics are measurable in experiment. Therefore, although Fisher information satisfies the Cramer-Rao bound \cite{amari,petz,cramer,fisher,amari1,fuji,matsumoto}, in quantum metrology or in precision measurement technique, we do not consider it as a useful measure of precision or uncertainty in parameter. Our metric does not avoid the non-uniqueness property but it can be experimentally measurable in experiment and can be a useful quantity in the precision measurement or metrology.

Starting from this new FS metric, we derive the square-root derivative \cite{luo1} Fisher information metric \cite{hels,hels1,yuen,barn,holevo} by imposing the monotonicity condition. We define the square-root derivative super-operator and show its relation with operator concave or convex functions. We also relate it with the existing logarithmic derivative super-operators so that we get the same Fisher information.

In this paper, we also generalized this newly developed FS metric for mixed states further from uniquely defined Fubini-Study metric \cite{pati,anandan,karol} for pure states by expressing the same purification of the mixed state in different bases. This generalization procedure is not unique. For each generalization procedure, we arrive at a set of metrics. We call them $\alpha$ metrics for mixed states. For $\alpha=1$, it reduces to the FS metric for mixed states. This newly defined $\alpha$ metric also satisfies the Cramer-Rao bound \cite{petz,amari,amari1,cramer,fisher,fuji,matsumoto}. Again, by imposing the monotonicity condition on this metric, we get monotonic $\alpha$ metric, which reduces to the Fisher information metric as in \cite{luo1} for $\alpha=1$.

The main motivation behind such generalization is the following. The dynamical phase \cite{pati} on the monotone Fisher information metric space is always zero. But we know that in any general evolution, the dynamical phase may be non-zero. The question then arises whether there exists monotone metric space on which dynamical phase in non-zero or not. The answer to this question is affirmative. The generalized monotone metric, i.e., the monotone $\alpha$ metric space is such a space on which dynamical phase is non-zero.

In the next section, we introduce the metric. In the third section, we prove the Cramer-Rao bound \cite{amari,petz,cramer,fisher,amari1,fuji,matsumoto}. Then, we study various properties of the metric. We relate it with the square-root derivative Fisher information \cite{luo1} and the quantum uncertainty as defined in \cite{luo}. In section {\bf V.}, we propose an experiment to measure the FS metric for mixed states. Next follows the generalization of the mixed state FS metric, from where we derive a set of monotone metrics called the monotonic $\alpha$ metrics.

%%%%%%%%%%%%%%%%%%%%%%%%%%%%%%%%%%%%%%%%%%%%%%%%%%%%%%%%%%%%%%%%%%%%%%%%%%%%%%

%%%%%%%%%%%%%%%%%%%%%%%%%%%%%%%%%%%%%%%%%%%%%%%%%%%%%%%%%%%%%%%%%%%%%%%%%%%%%%%%%
\section{The metric}
%%%%%%%%%%%%%%%%%%%%%%%%%%%%%%%%%%%%%%%%%%%%%%%%%%%%%%%%%%%%%%%%%%%%%%%%%%%%%
Let ${\cal H}_n$ denotes n dimensional Hilbert space and ${\cal L}({\cal H})$ is the set of linear
operators on ${\cal H}_n$. Let us consider a system $A$ with a mixed state $\rho_{A}(\boldsymbol{\theta})\in\mathcal{S(H}_{n})$ (the set of positive definite density operators on $\mathcal{H}_{n}$), where $\boldsymbol{\theta}$ is a d-tuple ($d=n^2-1$) defined as $\boldsymbol{\theta}=(\theta_1,\theta_2,....,\theta_d)$. From here on the subscript $A$ on $\rho$ has been dropped and should be understood as a state of the subsystem $A$. The distance between two pure quantum states is measured by the Fubini-Study distance \cite{pati,anandan,karol} which is nothing but a gauge invariant distance. Here, we give a similar distance between two mixed states. We demand that even if the system is in a mixed state, the purified version of the state in the enlarged Hilbert space, i.e., the system and the ancilla or environment must evolve under the equation of motion, which is gauge invariant. We purify the state by adding an ancilla $B$ of at least equal dimension. The purified state \cite{nielsen,karol} is given by
\begin{equation}\label{m}
\vert\Psi_{AB}(\boldsymbol{\theta})\rangle=(\sqrt{\rho(\boldsymbol{\theta})}V_{A}\otimes 
V_{B})\vert\alpha\rangle\in \cal{H}_{A}\otimes\cal{H}_{B},
\end{equation}
where $\vert\alpha\rangle=\underset{i}{\sum}\vert i^A i^B\rangle$ and $V_{A}$, $V_{B}$ are unitary operators on the subsystems $A$ and $B$ respectively. We know that $|d\psi\rangle=d\theta\frac{\partial|\psi\rangle}{\partial\theta}$ and that
\begin{equation}\label{c}
|d\psi_{\perp}\rangle=|d\psi\rangle-\frac{|\psi\rangle\langle\psi|}{\langle\psi|\psi\rangle}|d\psi\rangle.
\end{equation}
Eq. (\ref{c}) gives a differential form which does not distinguish two collinear vectors, i.e., $|\psi\rangle$ and $e^{i\theta}|\psi\rangle$ but $|d\psi\rangle$ does. But we want a distance on the projective Hilbert space where collinear vectors are taken to be the same vector. Therefore, we need $|d\psi_{\perp}\rangle$. 

Now, the angular variation of $|d\psi_{\perp}\rangle$ is
\begin{equation}\label{d}
|d\psi_{projec}\rangle=\frac{|d\psi_{\perp}\rangle}{\sqrt{\langle\psi|\psi\rangle}}=\frac{|d\psi\rangle}{\sqrt{\langle\psi|\psi\rangle}}-\frac{|\psi\rangle\langle\psi|}{\langle\psi|\psi\rangle^{3/2}}|d\psi\rangle.
\end{equation}
The FS metric \cite{pati,karol,anandan} on the Hilbert space is given by
\begin{equation}\label{e}
ds^{2}_{FS}=\langle d\psi_{projec}|d\psi_{projec}\rangle.
\end{equation}
The angular variation of the perpendicular component of the differential form in this case is given by
 \begin{equation}\label{n}
 |d\Psi_{AB_{projec}}(\boldsymbol{\theta})\rangle=d\theta_{i}(A^{i}_{\rho}-B^{i}_{\rho})|\alpha\rangle,
 \end{equation}
 where $A^{i}_{\rho}=(C^{i}_{\rho}V_{A}\otimes V_{B})$, $B^{i}_{\rho}=\vert\Psi_{AB}(\boldsymbol{\theta})\rangle\langle\Psi_{AB}(\boldsymbol{\theta})|(C^{i}_{\rho}V_{A}\otimes V_{B})$ with $C^{i}_{\rho}=\partial^{i}\sqrt{\rho(\boldsymbol{\theta})}$, the square root derivative.
 Therefore, the FS metric is given by
 \begin{eqnarray}\label{o}
&ds^{2}_{FS}&=\langle d\Psi_{AB_{projec}}(\boldsymbol{\theta})|d\Psi_{AB_{projec}}(\boldsymbol{\theta})\rangle\nonumber\\&=&d\theta_{i}d\theta_{j}[\langle\beta|(A^{i^\dagger}_{\rho}A^{j}_{\rho}-A^{i^\dagger}_{\rho}B^{j}_{\rho}-B^{i^\dagger}_{\rho}A^{j}_{\rho}+B^{i^\dagger}_{\rho}B^{j}_{\rho})|\alpha\rangle]\nonumber\\&=&d\theta_{i}d\theta_{j}[Tr(C^{i^{\dagger}}_{\rho}C^{j}_{\rho})-Tr(\sqrt{\rho(\boldsymbol{\theta})}C^{i^{\dagger} }_{\rho})Tr(\sqrt{\rho(\boldsymbol{\theta})}C^{j}_{\rho})].\nonumber\\
\end{eqnarray}
where $|\beta\rangle=\sum_{j}|j^{A}j^{B}\rangle$. On the third line of Eq. (\ref{o}), we also assumed the fact that $\langle i|j\rangle=\delta_{ij}$, i.e., they are on the same basis. Therefore, the quantum geometric tensor (QGT) \cite{cheng} is given by 
$g^{ij}_{\rho}$=$Tr(C^{i^{\dagger}}_{\rho}C^{j}_{\rho})-Tr(\sqrt{\rho
(\boldsymbol{\theta})}C^{i^{\dagger}}_{\rho})Tr(\sqrt{\rho
(\boldsymbol{\theta})}C^{j}_{\rho})=(\gamma^{ij}_{\rho}+\mathbf{i}\sigma^{ij}_{\rho})$. $ds^{2}_{FS}$ is the norm of $|d\Psi_{AB_{projec}}(\boldsymbol{\theta})\rangle$. Therefore, $g^{ij}_{\rho}$ must be Hermitian or in other words conjugate symmetric, i.e., $g^{ij}=(g^{ji})^{*}$. This implies that the real part of the metric tensor must be symmetric giving rise to a Riemannian structure and the imaginary part must be anti-symmetric giving rise to the  Fubini-Study symplectic structure. The quantity $d\theta_{i}d\theta_{j}\sigma^{ij}_{\rho}$ vanishes due to the symmetric and anti-symmetric combinations. Thus, the distance is given by $ds^{2}_{FS}=d\theta_{i}d\theta_{j}\gamma^{ij}_{\rho}$.
%\begin{eqnarray}\label{o}
%&&ds^{2}_{FS}=d\theta_{i}d\theta_{j}\gamma^{ij}_{\rho}\nonumber\\&&=d\theta_{i}
%d\theta_{j}Re[Tr(C^{i^{\dagger}}_{\rho}C^{j}_{\rho})-Tr(\sqrt{\rho_{A}(\boldsymbol{\theta})}C^{i^{\dagger}}_{\rho})Tr(\sqrt{\rho_{A} (\boldsymbol{\theta})}C^{j}_{\rho})].\nonumber\\
%\end{eqnarray}
Depending on the definition of the operator $C^{i}_{\rho}$, we can get various metric tensors. 

This is the most natural generalization of the Fubini-Study distance for mixed states. It is a $U(1)$ gauge invariant distance along any parameter $\boldsymbol{\theta}$ for mixed states unlike the FS metric for pure states, in which case the distance is along the unitary orbit only.
%
%We know that $Tr[\sqrt{\rho}C^{i}_{\rho}]$ is invariant under unitary transformation. So, in the basis where the density matrix is diagonal, we get $Tr[\sqrt{\rho}C^{i}_{\rho}]=\sum_{i,j}\sqrt{\lambda_{i}}[C^{i}_{\rho}]_{ij}$ which is a purely imaginary number because we know that $Tr[\sqrt{\rho}C^{i}_{\rho}]+Tr[\sqrt{\rho}C^{i^{\dagger}}_{\rho}]=0$ (Using $Tr[\rho]=1$). So, the matrix $C^{i}_{\rho}$ must be a purely imaginary matrix.

%%%%%%%%%%%%%%%%%%%%%%%%%%%%%%%%%%%%%%%%%%%%%%%%%%%%%%%%%%%%%%%%%%%%%%%%%%%%%%%%%%%%%
\section{Cramer-Rao bound}
%%%%%%%%%%%%%%%%%%%%%%%%%%%%%%%%%%%%%%%%%%%%%%%%%%%%%%%%%%%%%%%%%%%%%%%%%%%%%%%%%%%%%%
It is well known that the Quantum Fisher Information for pure states reduces to the Fubini-Study distance. Thus, the Fubini-Study distance \cite{karol,anandan,pati} satisfies the quantum Cramer-Rao bound \cite{fuji,petz,matsumoto,amari,amari1,cramer,fisher,watanabe}. We derived our metric starting from the Fubini-Study distance. Therefore, it must satisfy the quantum version of the Cramer-Rao Bound as
\begin{equation}\label{p}
\Delta\lambda^2\geq\frac{1}{\gamma_{\rho}(\partial_{\lambda}\rho,\partial_{\lambda}\rho)}.
\end{equation}
Note that the quantity $\gamma_{\rho}(\partial_{\lambda}\rho,\partial_{\lambda}\rho)$ is neither the Fisher information nor monotonic. 

An important point to note is that the bound given here is experimentally measurable in the interference experiment by observing the phase shift of the interference pattern \cite{deba}. In our paper, we have proposed an experiment to measure the infinitesimal distance Eq. (\ref{o}) in section V.

%%%%%%%%%%%%%%%%%%%%%%%%%%%%%%%%%%%%%%%%%%%%%%%%%%%%%%%%%%%%%%%%%%%%%%%%%%%%%
\section{Properties of the metric}
%%%%%%%%%%%%%%%%%%%%%%%%%%%%%%%%%%%%%%%%%%%%%%%%%%%%%%%%%%%%%%%%%%%%%%%%%%%%%
 At first let us write the generalized Fubini-Study metric tensor $\gamma^{ij}_{\rho}$ in terms of tangents of the tangent space $T_{\rho}$ (set of traceless hermitian operators) on $S(\mathcal{H}_{n})$ \cite{petz,watanabe}. To do so, we will replace indices of the metric i, j with $A, B\in T_{\rho}$ respectively as
$\gamma^{ij}_{\rho}\rightarrow\gamma_{\rho}(A,B)$.
Let us now write the Square root derivative $C^{i}_{\rho}$ in terms of operators on tangent space as $C^{i}_{\rho}\rightarrow \mathbf{K}^{-1}_{\rho}(A)$, where $\mathbf{K}_{\rho}$ is a positive super-operator \cite{petz,watanabe}. So, the metric in terms of super-operators on tangent space is given by
$\gamma_{\rho}(A,B)=Re[\langle \mathbf{K}^{-1}_{\rho}(A), \mathbf{K}^{-1}_{\rho}(B)\rangle-\langle \mathbf{K}^{-1}_{\rho}(A),\sqrt{\rho}\rangle\langle\sqrt{\rho},\mathbf{K}^{-1}_{\rho}(B)\rangle]$,
where $\langle .,. \rangle$ is the Hilbert-Schmidt inner product. Here we impose an extra condition that it satisfies the monotonicity property which is an important requirement for a metric in quantum information theory. We say that the metric is monotone if and only if $\gamma_{{\cal E}(\rho)}({\cal E}(A),{\cal E}(A))\leq \gamma_{\rho}(A,A)$ for all $A\in T_{\rho}$, for all $\rho\in S(\mathcal{H}_{n})$ and for all completely positive trace preserving (CPTP) map ${\cal E}:S(\mathcal{H}_{n})\rightarrow S(\mathcal{H}_{m})$. 
%Equality holds for an operation ${\cal E}$ which has ${\cal E'}$ such that ${\cal E}{\cal E}(X)=X$ for all $X\in \mathcal{L(H}_{n})$ \cite{petz}.
 Monotonicity condition also includes the covariance of the metric tensor under unitary operation, i.e., $\gamma_{U\rho U^{\dagger}}(UAU^{\dagger},UBU^{\dagger})=\gamma_{\rho}(A,B)$. 
 
We denote the first term of $\gamma_{\rho}$ as F, second term as S and under CPTP map they become $F_{\cal E}$ and $S_{\cal E}$ respectively. So, the metric will be monotonic if $F_{\cal E}-S_{\cal E}\leq F-S$. To show the monotonicity property, we consider all those positive super-operators, which satisfies the inequality ${\cal E}^{\dagger}\mathbf{K}^{-1}_{{\cal E}(\rho)}{\cal E}\geq \mathbf{K}^{-1}_{\rho}$ or, ${\cal E}^{\dagger}\mathbf{K}^{-1}_{{\cal E}(\rho)}{\cal E}\leq \mathbf{K}^{-1}_{\rho}$. We show below that both ${\cal E}^{\dagger}\mathbf{K}^{-1}_{{\cal E}(\rho)}{\cal E}\geq \mathbf{K}^{-1}_{\rho}$ and ${\cal E}^{\dagger}\mathbf{K}^{-1}_{{\cal E}(\rho)}{\cal E}\leq \mathbf{K}^{-1}_{\rho}$ correspond to either operator concave or operator convex functions with certain properties. Using such properties and the unitary invariance of the metric, we show that the second term vanishes for both kinds of super-operators.
% \begin{eqnarray}\label{17}
% &Tr[\sqrt{\rho}{\cal E}^{\dagger}K^{-1}_{{\cal E}(\rho)}{\cal E} A]\leq Tr[\sqrt{\rho} K^{-1}_{\rho} A]\nonumber\\  \implies &Tr[A{\cal E}^{\dagger}K^{-1}_{{\cal E}(\rho)}\sqrt{{\cal E}(\rho)}]Tr[\sqrt{{\cal E}(\rho)}K^{-1}_{{\cal E}(\rho)}{\cal E} A]\leq Tr[A K^{-1}_{\rho}\sqrt{\rho} ]Tr[\sqrt{\rho} K^{-1}_{\rho} A]\nonumber\\ \implies &\langle K^{-1}_{{\cal E}(\rho)}{\cal E}(A),\sqrt{{\cal E}(\rho)}\rangle\langle\sqrt{{\cal E}(\rho)},K^{-1}_{{\cal E}(\rho)}{\cal E}(A)\rangle\leq\langle K^{-1}_{\rho}(A),\sqrt{\rho}\rangle \langle\sqrt{\rho},K^{-1}_{\rho}(A)\rangle.\nonumber\\
% \end{eqnarray}
% When $A\leq 0$,
% \begin{eqnarray}
% &Tr[\sqrt{\rho}{\cal E}^{\dagger}K^{-1}_{{\cal E}(\rho)}{\cal E} A]\geq Tr[\sqrt{\rho} K^{-1}_{\rho} A]\nonumber\\ \implies & [\sqrt{\rho}{\cal E}^{\dagger}K^{-1}_{{\cal E}(\rho)}{\cal E} A]^{\dagger}Tr[\sqrt{\rho}{\cal E}^{\dagger}K^{-1}_{{\cal E}(\rho)}{\cal E} A]\leq [\sqrt{\rho} K^{-1}_{\rho} A]^{\dagger}Tr[\sqrt{\rho} K^{-1}_{\rho} A],\nonumber\\
% \end{eqnarray}
% which again implies the same inequality.

Now let us find out the condition on the super-operator for the metric to be monotonic. As the second term vanishes, we get the monotonicity condition as $F_{\cal E}\leq F$, i.e., $\langle \mathbf{K}^{-1}_{{\cal E}(\rho)}{\cal E}(A), \mathbf{K}^{-1}_{{\cal E}(\rho)}{\cal E}(B)\rangle\leq\langle \mathbf{K}^{-1}_{\rho}(A), \mathbf{K}^{-1}_{\rho}(B)\rangle$, which is satisfied if and only if
\begin{eqnarray}\label{18}
&& {\cal E}^{\dagger}\mathbf{K}^{-2}_{{\cal E}(\rho)}{\cal E}\leq \mathbf{K}^{-2}_{\rho}\Leftrightarrow{\cal E}\mathbf{K}^{2}_{\rho}{\cal E}^{\dagger}\leq \mathbf{K}^{2}_{{\cal E}(\rho)}.
\end{eqnarray}
This is the monotonicity condition of the metric when super-operators satisfies either ${\cal E}^{\dagger}\mathbf{K}^{-1}_{{\cal E}(\rho)}{\cal E}\geq \mathbf{K}^{-1}_{\rho}$ or ${\cal E}^{\dagger}\mathbf{K}^{-1}_{{\cal E}(\rho)}{\cal E}\leq \mathbf{K}^{-1}_{\rho}$. 

It was shown that the super-operators for which the quantum logarithmic Fisher information is monotone, can always be expressed in terms of operator concave functions \citep{petz,watanabe}. But Below we show that not only operator concave functions but also other convex functions provide super-operators which give rise to the square-root Fisher information. Here we show in the following theorem that we can always generate monotone metrics from known monotone metrics.

{\bf Theorem.---} Let $\mathbf{K}^{(1)}_{\rho}$ and $\mathbf{K}^{(2)}_{\rho}$ be two positive super-operators corresponding to monotone metrics satisfying the monotonicity condition ${\cal E}^{\dagger}\mathbf{K}^{-1}_{{\cal E}(\rho)}{\cal E}\geq \mathbf{K}^{-1}_{\rho}$. Then for any operator $\sigma$-mean \cite{supp} $\sigma$ and operator mean $\tau$, $\mathbf{K}_{\rho}=\mathbf{K}^{(1)}_{\rho}\sigma \mathbf{K}^{(2)}_{\rho}$ such that $\mathbf{K}^2_{\rho}=\mathbf{K}^{(1)^2}_{\rho}\tau \mathbf{K}^{(2)^2}_{\rho}$, gives another monotone metric.

$Proof.${---}  Using the monotonicity property and the transformer inequality in \cite{supp} we prove the first inequality of the monotonicity condition as
\begin{eqnarray}\label{19}
{\cal E}\mathbf{K}_{\rho}{\cal E}^{\dagger}&=&{\cal E}\mathbf{K}^{(1)}_{\rho}\sigma \mathbf{K}^{(2)}_{\rho}{\cal E}^{\dagger}\nonumber\\&\geq &{\cal E}\mathbf{K}^{(1)}_{\rho}{\cal E}^{\dagger}\sigma{\cal E} \mathbf{K}^{(2)}_{\rho}{\cal E}^{\dagger}\nonumber\\&\geq &\mathbf{K}^{(1)}_{\cal E(\rho)}\sigma \mathbf{K}^{(2)}_{\cal E(\rho)}=\mathbf{K}_{\cal E(\rho)}.
\end{eqnarray}
To prove the second inequality, let us start with
\begin{eqnarray}\label{20}
{\cal E}\mathbf{K}^{2}_{\rho}{\cal E}^{\dagger}&=&{\cal E}(\mathbf{K}^{(1)^2}_{\rho}\tau \mathbf{K}^{(2)^2}_{\rho}){\cal E}^{\dagger}\nonumber\\&\leq &{\cal E}\mathbf{K}^{(1)^2}_{\rho}{\cal E}^{\dagger}\tau{\cal E} \mathbf{K}^{(2)^2}_{\rho}{\cal E}^{\dagger}\nonumber\\&\leq &\mathbf{K}^{(1)^2}_{\cal E(\rho)}\tau \mathbf{K}^{(2)^2}_{\cal E(\rho)}=\mathbf{K}^2_{\cal E(\rho)}.
\end{eqnarray}
Therefore, $\mathbf{K}_{\rho}$ must provide a monotone metric.

{\bf Lemma.---} For all operator convex function $f:
\mathbb{R}^{+2}\rightarrow\mathbb{R}^{+}$ and operator monotone function $g:
\mathbb{R}^{+2}\rightarrow\mathbb{R}^{+}$, super-operators $\mathbf{K}_{\rho}
=f(\mathbf{P}_{\rho}\mathbf{T}^{-1}_{\rho})\mathbf{T}_{\rho}$ such that $\mathbf{K}^2_{\rho}
=g(\mathbf{P}^2_{\rho}\mathbf{T}^{-2}_{\rho})\mathbf{T}^2_{\rho}$ \cite{petz,watanabe}, where $\mathbf{P}_{\rho}$ and $\mathbf{T}_{\rho}$ satisfy both the monotonicity inequalities ${\cal E}^{\dagger}\mathbf{K}^{-1}_{{\cal E}(\rho)}{\cal E}\geq \mathbf{K}^{-1}_{\rho}$, ${\cal E}^{\dagger}\mathbf{K}^{-2}_{{\cal E}(\rho)}{\cal E}\leq \mathbf{K}^{-2}_{\rho}$ and commute with each-other, determine the monotone metric. 

$Proof.${---}  The super-operators $\mathbf{P}_{\rho}$ and $\mathbf{T}_{\rho}$ satisfy both the monotonicity inequalities. So, from the above theorem, we know that all the super-operators of the form $\mathbf{K}_{\rho}=\mathbf{P}_{\rho}\sigma\mathbf{T}_{\rho}$ such that $\mathbf{K}^2_{\rho}=\mathbf{P}^2_{\rho}\tau\mathbf{T}^2_{\rho}$for any operator $\sigma$-means $\sigma$ and operator means $\tau$, satisfy the monotonicity condition and give monotonic metrics. Again we know that every operator $\sigma$-means can be expressed by operator convex function $f$ \cite{supp} and takes a form as stated above when $\mathbf{P}_{\rho}$ and $\mathbf{T}_{\rho}$ commute with each-other \cite{supp}. If the super-operator $\mathbf{K}_{\rho}$ satisfies ${\cal E}^{\dagger}\mathbf{K}^{-1}_{{\cal E}(\rho)}{\cal E}\leq \mathbf{K}^{-1}_{\rho}$ , it can be easily shown that $f$ is operator concave function \cite{petz} and any super-operator of the form $\mathbf{K}_{\rho}=\mathbf{K}_{\rho}^{(1)}\sigma \mathbf{K}_{\rho}^{(2)}$, where $\sigma$ is operator mean \cite{petz,watanabe}, gives monotone metric.

We are already familiar with logarithmic derivative Fisher information metric and corresponding super operators $K^{l}_{\rho}$\cite{petz,luo1}. Here superscript $l$ denotes super-operator corresponding to logarithmic derivative. Similarly, we denote a super-operator of above class corresponding to square-root derivative as $K^{s}_{\rho}$\cite{petz,luo1}. Suppose they both produces the same Fisher information. Then if $K^{l}_{\rho}=g(L^{l}_{\rho}R^{l-1}_{\rho})R^{l}_{\rho}$, where $g$ is a positive operator concave function \cite{petz}, it is easy to show that $K^{s}_{\rho}=\sqrt{g(L^{s2}_{\rho}R^{s-2}_{\rho})R^{s2}_{\rho}}$, when $L_{\rho}$ and $R_{\rho}$ commute with each-other and are right and left square-root or logarithmic derivatives respectively. 

 We know that the monotonicity of the distance implies the unitary invariance. Therefore, from the second term we get
\begin{eqnarray}\label{21}
Tr(\sqrt{\rho(\mathbf{\theta})}C_{\rho}^{i\dagger})&=&\sum_{i,j}\sqrt{\rho(\mathbf
{\theta})}_{ij}[C_{\rho}^{i\dagger}]_{ji}\nonumber\\&=&\sum_{i}\sqrt{\lambda_{i}}\frac{d\rho_{ii}}{\sqrt{\lambda_{i}}}=0,
\end{eqnarray}
where the last line is due to the fact that the quantity is unitarily invariant (we also considered $[C_{\rho}^{i\dagger}]_{ij}=\frac{d\rho_{ij}}{f(\lambda_{i}^{\frac{1}{2}}\lambda_{j}^{-\frac{1}{2}})\lambda_{j}^{\frac{1}{2}}}$, where $f$ is either operator concave or convex and $f(1)=1$). 
Therefore, the second term or the dynamical phase \cite{pati}, i.e., $i\langle\psi_{AB}|d\psi_{AB}\rangle=i\text{Tr}(C^{i}_{\rho}\sqrt{\rho})$ is zero if we impose the monotonicity condition on the generalized Fubini-Study metric. It reduces to square-root derivative Fisher information \cite{luo1}, when the square-root derivative satisfies the monotonicity conditions mentioned above. Therefore, on the Fisher information metric space, dynamical phase is always zero. 

In the next subsections, we study the properties of this metric when the state evolves under unitary or completely positive trace preserving (CPTP) evolutions.
\subsection{METRIC UNDER UNITARY EVOLUTION}
%%%%%%%%%%%%%%%%%%%%%%%%%%%%%%%%%%%%%%%%%%%%%%%%%%%%%%%%%%%%%%%%%%%%%%%%%
Suppose, a system with a state $\rho$ evolves under $U_{A}=e^{iH_{A}t}$. If we consider a purification \cite{karol,nielsen} of the state $\rho$ in the extended Hilbert space as $\vert\Psi_{AB}\rangle=(\sqrt{\rho}V_{A}\otimes 
V_{B})\vert\alpha\rangle\in \cal{H}_{A}\otimes\cal{H}_{B}$, the state at time t, must be $\vert\Psi_{AB}(t)\rangle=(\sqrt{\rho(t)}V_{A}\otimes 
V_{B})\vert\alpha\rangle=(U_{A}\sqrt{\rho}U_{A}^{\dagger}V_{A}\otimes 
V_{B})\vert\alpha\rangle$. Therefore, it is easy to show that the infinitesimal distance Eq. (\ref{o}),
\begin{eqnarray}\label{uni}
ds^{2}_{FS}=-dt^2[\text{Tr}[\sqrt{\rho},H_{A}]^2],
\end{eqnarray}
which is nothing but the quantum uncertainty as defined in \cite{luo}. Under unitary evolution, the second term of the metric (\ref{o}), i.e., the dynamical phase, $i\langle\psi_{AB}|\psi_{AB}\rangle$ is zero. But this is not the case for CPTP evolutions as shown in the next subsection.

\subsection{METRIC UNDER CPTP EVOLUTION}
%%%%%%%%%%%%%%%%%%%%%%%%%%%%%%%%%%%%%%%%%%%%%%%%%%%%%%%%%%%%%%%%%%
Suppose, a system with a state $\rho$ evolves under a CPTP map ${\cal E}$, whose Kraus operator representation is given by a set of operators $\{A_{i}\}$ such that $\sum_{i}A_{i}^{\dagger}A_{i}=I$. If we consider a purification \cite{karol,nielsen} of the state $\rho$ in the extended Hilbert space as $\vert\Psi_{AB}\rangle=(\sqrt{\rho}V_{A}\otimes 
V_{B})\vert\alpha\rangle\in \cal{H}_{A}\otimes\cal{H}_{B}$, the state at time t, must be $\vert\Psi_{AB}(t)\rangle=(\sqrt{\rho(t)}V_{A}\otimes 
V_{B})\vert\alpha\rangle=(\sum_{i}A_{i}\sqrt{\rho}A_{i}^{\dagger}V_{A}\otimes 
V_{B})\vert\alpha\rangle$, where $|\alpha\rangle=\sum_{j}|j^{A}j^{B}\rangle$. Therefor, it is again easy to show that the infinitesimal distance Eq. (\ref{o}),
\begin{eqnarray}\label{cptp}
ds^2_{FS}&=&\sum_{ij}\text{Tr}[(\dot{A}_{i}\sqrt{\rho}A_{i}^{\dagger}+A_{i}\sqrt{\rho}\dot{A}_{i}^{\dagger})(\dot{A}_{j}\sqrt{\rho}A_{j}^{\dagger}+A_{j}\sqrt{\rho}\dot{A}_{j}^{\dagger})]
\nonumber\\&-&|\sum_{i}\text{Tr}(\sqrt{\rho}(\dot{A}_{i}\sqrt{\rho}A_{i}^{\dagger}+A_{i}\sqrt{\rho}\dot{A}_{i}^{\dagger}))|^2.
\end{eqnarray}
From this quantity it is not clear whether it also gives the quantum uncertainty under CPTP evolutions or not. To show that it indeed captures the quantum uncertainty due to CPTP evolutions or gives the quantum part of the $U(1)$ gauge invariant distance along the CPTP evolution orbit ${\cal E}$, we need to show that it satisfies all the conditions listed in \cite{luo,wigner}. To do that we do not directly use the Eq. (\ref{cptp}). Instead, we use the fact that this CPTP evolution can always be represented as a unitary evolution in an extended Hilbert space via Stinespring's dilation theorem. We start from the Eq. (\ref{uni}) and consider the state $\rho\otimes|\nu\rangle_{B}\langle\nu|$ evolves under $U_{AB}=e^{iH_{AB}t}$. Then, the infinitesimal FS metric from the Eq. (\ref{uni}) reduces to
\begin{eqnarray}\label{cptp1}
ds^2_{FS}&=&-dt^{2}\text{Tr}[\sqrt{\rho}\otimes|\nu\rangle_{B}\langle\nu|, H_{AB}]^2\nonumber\\&=&2dt^{2}\text{Tr}[\tilde{H^{2}_{A}}\rho -\tilde{H_{A}}\sqrt{\rho}\tilde{H_{A}}\sqrt{\rho}],
\end{eqnarray}
where $\tilde{H_{A}}=_{B}\langle\nu|H_{AB}|\nu\rangle_{B}$ and $\tilde{H^{2}_{A}}=_{B}\langle\nu|H^{2}_{AB}|\nu\rangle_{B}$. This quantity is nothing but the quantity given in Eq. (\ref{cptp}). It is easy from the Eq. (\ref{cptp1}) that the infinitesimal FS distance for mixed states captures the quantum part of the uncertainty in the evolution operator ($H_{AB}$ here) because the quantity given in Eq. (\ref{cptp1}) is convex with respect to $\rho$ and is zero whenever $[\rho\otimes|\nu\rangle_{B}\langle\nu|,H_{AB}]=0$.

Therefore, it justifies our claim that the infinitesimal distance given in Eq. (\ref{o}) is the most natural generalization of the FS metric for mixed states.   This distance, like the FS metric for pure states is $U(1)$ gauge invariant and captures only the quantum uncertainty in the evolution Hamiltonian. The distance becomes zero whenever the evolution Hamiltonian commutes with the state.
%In the next section, we shall drop one of our assumptions and derive the Fubini-study metric again for mixed states. And by imposing the monotonicity condition and unitary invariance, we generalize the Fisher information from this new form of Fubini-Study metric.
%%%%%%%%%%%%%%%%%%%%%%%%%%%%%%%%%%%%%%%%%%%%%%%%%%%%%%%%%%%%%%%%%%%%%%%%%%%%%%%%
%%%%%%%%%%%%%%%%%%%%%%%%%%%%%%%%%%%%%%%%%%%%%%%%%%%%%%%%%%%%%%%%%%%%%%%%%%
\section{Experimental Proposal}
%%%%%%%%%%%%%%%%%%%%%%%%%%%%%%%%%%%%%%%%%%%%%%%%%%%%%%%%%%%%%%%%%%%%%%%%%%%%
In this section, we consider the simplest of all evolutions, the unitary evolution. The state of the system $\rho(0)$ evolves to $\rho(t)$ under a unitary operator $U_{A}=e^{iH_{A}t}$. Under such evolutions (unitary), the infinitesimal distance Eq. (\ref{o}) is nothing but the quantum uncertainty as defined by S. Luo in \citep{luo}. This quantum uncertainty can not be measured in the interference experiment. But we know that the variance of an evolving operator, i.e., Hamiltonian can be measured. Our aim here is to equate our infinitesimal distance with such an uncertainty or variance of an operator. We suppose that the purification \cite{karol,nielsen} of the state $\rho(0)$ in the extended Hilbert space $|\psi_{AB}\rangle=\sqrt{\rho(0)}|\alpha\rangle$ evolves to $|\psi_{AB}(t)\rangle=\sqrt{\rho(t)}|\alpha\rangle$ under an another unitary operator $U_{AB}=e^{iH_{AB}t}$, i.e.,
\begin{equation}\label{ep}
e^{iH_{AB}t}|\psi_{AB}(0)\rangle=\sqrt{\rho(t)}|\alpha\rangle.
\end{equation} 
Therefore, we need to find out $H_{AB}$ and measure its variance. Comparing the Taylor expansions of both the sides of the Eq. (\ref{ep}), we get
\begin{eqnarray}\label{p1}
iH_{AB}\sqrt{\rho(0)}&=&\partial_{t}\sqrt{\rho(t)}|_{t\rightarrow 0},\nonumber\\H_{AB}^2\sqrt{\rho(0)}&=&
-\partial_{t}^2\sqrt{\rho(t)}|_{t\rightarrow 0}
\end{eqnarray}
and so on. To get the infinitesimal distance in the interference experiment, it is enough to find out an operator $H_{AB}$, which satisfies at least the first equation in Eq. (\ref{p1}). It is because the infinitesimal distance Eq. (\ref{o}), only depends on the first derivative of $\sqrt{\rho}$. Now, the right hand side of the first equation in Eq. (\ref{p1}) is known because under the unitary operator $U_{A}$, we can easily show that $\partial_{t}\sqrt{\rho(t)}=i[\sqrt{\rho(t)},H_{A}]$. As we know both $\rho$ and $H_{A}$, we can easily calculate one such $H_{AB}$, which satisfies the first relation in Eq. (\ref{p1}). In the interference experiment, we can easily measure the variance of $H_{AB}$ in one of the purified states, $|\psi_{AB}\rangle$ by measuring the shift in the interference pattern \cite{deba}. This quantity is nothing but our infinitesimal FS distance due to the evolution of the state $\rho(0)$ under $U_{A}$. In other words, we have connected the quantum uncertainty as defined in \cite{luo} with a measurable quantity. The infinitesimal distance for CPTP evolutions can also be measured as the evolution can be written as a unitary evolution in the extended Hilbert space \cite{deba,karol} via stinespring's dilation theorem.

In the previous section, we have shown that the dynamical phase is zero on the monotone metric space. Therefore, the next logical question to ask is that whether there exists any monotone metric space on which the dynamical phase is non-zero, i.e., the dynamical developed during the evolution of the state along the geodesics of monotone metric is non-zero. In the next section, we answer to this question in the affirmative sense.

%%%%%%%%%%%%%%%%%%%%%%%%%%%%%%%%%%%%%%%%%%%%%%%%%%%%%%%%%%%%%%%%%%%%%%%%%%%%%%%
\section{Generalization of Fisher information}
%%%%%%%%%%%%%%%%%%%%%%%%%%%%%%%%%%%%%%%%%%%%%%%%%%%%%%%%%%%%%%%%%%%%%%%%%%%%%%%%%%%%%
In the last sections, we showed that the dynamical phase \cite{pati} on the monotone metric space is always zero. But we know that it may be non-zero in any general evolution. One important motivation to generalize the Fisher information in this section is to get monotone metric on which the dynamical phase is non-zero. To get that we dropped a few assumptions as described below.

In the Eq. (\ref{o}), we considered that $|\alpha\rangle=\sum_{i}|i^{A}i^{B}\rangle$, $|\beta\rangle=\sum_{j}|j^{A}j^{B}\rangle$ and assumed $\langle i|j\rangle=\delta_{ij}$. In this section, we drop out such assumption. Instead, we consider $|\alpha\rangle=\sum_{i}|A^{A}_{i}A^{B}_{i}\rangle$ and $|\beta\rangle=\sum_{i}|B^{A}_{i}B^{B}_{i}\rangle$ such that $\langle A_{i}|B_{j}\rangle\neq\delta_{ij}$. The motivation behind such consideration is following. Suppose a state $\rho(t)$ at time $t$ evolves to $\rho(t+\Delta t)$ at time $t+\Delta t$. A purification \cite{karol,nielsen} of $\rho(t)$, in principle, may evolve to any of the purifications \cite{karol,nielsen} of $\rho(t+\Delta t)$, i.e., $\vert\Psi_{AB}(\boldsymbol{\theta})\rangle=(\sqrt{\rho(\boldsymbol{\theta})}V_{A}\otimes 
V_{B})\vert\alpha\rangle$ may evolve to $\vert\tilde{\Psi}_{AB}(\boldsymbol{\theta}+\Delta\boldsymbol{\theta})\rangle=(\sqrt{\rho(\boldsymbol
{\theta}+\Delta\boldsymbol{\theta})}V_{A}\otimes 
V_{B})\vert\beta\rangle$, where $|\alpha\rangle$ and $|\beta\rangle$ are on two different bases and tilde sign is to denote that they are expressed on two different bases. Therefore, we can now define the differential of the state either by differentiating the initial state $\vert\Psi_{AB}(\boldsymbol{\theta})\rangle$ or by Taylor expanding the final state $\vert\tilde{\Psi}_{AB}(\boldsymbol{\theta}+\Delta\boldsymbol{\theta})\rangle$. Following these two methods, we get $|d\psi\rangle$ and $|d\tilde{\psi}\rangle$. As both of these quantities are the same expressed in two different bases, we can now use then interchangeably in the definition of Fubini-Study metric. In the limit $\Delta\boldsymbol{\theta}\rightarrow 0$, $\vert\tilde{\Psi}_{AB}(\boldsymbol{\theta}+\Delta\boldsymbol{\theta})\rangle$ must reduce to $\vert\Psi_{AB}(\boldsymbol{\theta})\rangle$, i.e., $\langle\tilde{\Psi}_{AB}
({\boldsymbol\theta})|\Psi_{AB}({\boldsymbol\theta})\rangle=1$. It is easy to 
show that this condition implies $\langle A_{i}|B_{k}\rangle\langle A_{i}|B_{j}\rangle=\{\Lambda_{\tilde{\rho}_{ai}}^{\alpha}|\alpha=2,3...n\}$ (as $\langle A_{i}|B_{j}\rangle\neq\delta_{ij}$) (see \cite{supp}), i.e., at each instant of time, the relation between the bases $\{B_{i}\}$ and $\{A_{i}\}$ depends on the instantaneous states. Therefore,  $|\tilde{\Psi}_{AB}({\boldsymbol\theta})\rangle$ and $|\Psi_{AB}({\boldsymbol\theta})\rangle$ are actually the same state expressed in two different bases. To derive the $\alpha$ metric, we use here the relation between the bases as given in the appendix B \citep{supp}.

We know that the Fubini-Study metric \cite{pati,karol,anandan} is given by 
\begin{eqnarray}\label{22}
ds^{2}_{FS}=\langle d\Psi|d\Psi\rangle-|\langle\Psi|d\Psi\rangle|^2.
\end{eqnarray}
We show that there is no unique way to generalize Fubini-Study metric for mixed states and Fisher information metric. Here, we generalize the Fubini-Study metric by replacing $|\Psi_{AB}({\boldsymbol\theta})\rangle$ by $|\tilde{\Psi}_{AB}({\boldsymbol\theta})\rangle$ in the following way
\begin{equation}\label{23}
ds^{2}_{FS}=\langle d\tilde{\Psi}|d\Psi\rangle-|\langle\tilde{d\Psi}|\Psi\rangle|^2.
\end{equation}
We define that $ |\Psi_{AB}(\boldsymbol{\theta})\rangle=A_{\rho}|\alpha\rangle$,  $ |\tilde{\Psi}_{AB}(\boldsymbol{\theta})\rangle=A_{\rho}|\beta\rangle$, where $A_{\rho}=\sqrt{\rho(\boldsymbol{\theta})}V_{A}\otimes 
V_{B}$, and know that $ |d\Psi_{AB}(\boldsymbol{\theta})\rangle=d\theta_{i}A^{i}_{\rho}|\alpha\rangle$, $ |d\tilde{\Psi}_{AB}(\boldsymbol{\theta})\rangle=d\theta_{i}A^{i}_{\rho}|\beta\rangle$, such that $A^{i}_{\rho}=(C^{i}_{\rho}V_{A}\otimes V_{B})$. Therefore, the generalized Fubini-Study metric using the Einstein's summation convention is given by
\begin{eqnarray}\label{24}
&&ds^{2}_{FS}=\langle\beta|A^{i\dagger}_{\rho}A^{i}_{\rho}|\alpha\rangle-|\langle\beta|A'^{\dagger}A|\alpha\rangle|^2\nonumber\\&=&G^{cd}_{\rho}(\alpha)d\theta_{c}d\theta_{d},
\end{eqnarray} 
where $G^{cd}_{\rho}(\alpha)$ is the generalized Fubini-Study metric-tensor and is given by $G^{cd}_{\rho}(\alpha)=[\frac{\text{Tr}(\rho^{\alpha-1}C^{c\dagger}_{\rho}C^{d}_{\rho})}{\text{Tr}(\rho^{\alpha})}-\frac{\text{Tr}(\rho^{\alpha-\frac{1}{2}}C^{c\dagger}_{\rho})}{\text{Tr}(\rho^{\alpha})}\frac{\text{Tr}(\rho^{\alpha-\frac{1}{2}}C^{d}_{\rho})}{\text{Tr}(\rho^{\alpha})}]$ (using the form of $\Lambda^{\alpha}_{\rho_{jk}}$ given in appendix B \citep{supp}). This reduces to the metric tensor given by Eq. (\ref{o}) for $\alpha=1$. We show here another way to generalize Fubini-Study metric below. For this, we define the metric as
\begin{eqnarray}\label{25}
ds^{2}_{FS}=\langle d\tilde{\Psi}_{AB_{projec}}(\boldsymbol{\theta})|d\Psi_{AB_{projec}}(\boldsymbol{\theta})\rangle.
\end{eqnarray}

The angular variation of the perpendicular component of the differential form in this case is given by
 \begin{eqnarray}\label{26}
 |d\Psi_{AB_{projec}}(\boldsymbol{\theta})\rangle=d\theta_{i}(A^{i}_{\rho}-B^{i}_{\rho})|\alpha\rangle,\\
  |d\tilde{\Psi}_{AB_{projec}}(\boldsymbol{\theta})\rangle=d\theta_{i}(A^{i}_{\rho}-\tilde{B}^{i}_{\rho})|\beta\rangle,
 \end{eqnarray}
 where $A^{i}_{\rho}=(C^{i}_{\rho}V_{A}\otimes V_{B})$, $B^{i}_{\rho}=\vert\Psi_{AB}(\boldsymbol{\theta})\rangle\langle\Psi_{AB}(\boldsymbol{\theta})|(C^{i}_{\rho}V_{A}\otimes V_{B})$ and $\tilde{B}^{i}_{\rho}=\vert\tilde{\Psi}_{AB}(\boldsymbol{\theta})\rangle\langle\tilde{\Psi}_{AB}(\boldsymbol{\theta})|(C^{i}_{\rho}V_{A}\otimes V_{B})$ with $C^{i}_{\rho}=\partial^{i}\sqrt{\rho_{A}(\boldsymbol{\theta})}$, the square root derivative.
The Fubiny-Study metric is given by
 \begin{eqnarray}\label{27}
&&ds^{2}_{FS}(\alpha)=\langle d\tilde{\Psi}_{AB_{projec}}(\boldsymbol{\theta})|d\Psi_{AB_{projec}}(\boldsymbol{\theta})\rangle\nonumber\\&=&d\theta_{c}d\theta_{d}[\langle\beta|(A^{c^\dagger}_{\rho}A^{d}_{\rho}-A^{c^\dagger}_{\rho}B^{d}_{\rho}-\tilde{B}^{c^\dagger}_{\rho}A^{d}_{\rho}+
\tilde{B}^{c^\dagger}_{\rho}B^{d}_{\rho})|
\alpha\rangle]\nonumber\\&=&d\theta_{c}d\theta_{d}[\frac{\text{Tr}(
\rho^{\alpha-1}C^{c^{\dagger}}_{\rho}C^{d}_{\rho})}{\text{Tr}(\rho^{\alpha})}-\frac{\text{Tr}
(\rho^{\alpha-\frac{1}{2}}C_{\rho}^{c\dagger})}{\text{Tr}(\rho^{\alpha})} \text{Tr}(\sqrt{\rho}C_{\rho}^{d})\nonumber\\&-&\frac{\text{Tr}(\rho^{\alpha-\frac{1}{2}}C_{\rho}^{d})}{\text{Tr}(\rho^{\alpha})} 
\text{Tr}(\sqrt{\rho}C_{\rho}^{c\dagger})+\text{Tr}(\sqrt{\rho}C^{c^{\dagger} 
}_{\rho})\text{Tr}(\sqrt{\rho}C^{d}_{\rho})],\nonumber\\
\end{eqnarray}
where we considered Einstein's summation convention. Therefore, the quantum geometric tensor (QGT) \cite{cheng} is given by 
$\tilde{G}^{cd}_{\rho}(\alpha)$=$\frac{\text{Tr}(
\rho^{\alpha-1}C^{c^{\dagger}}_{\rho}C^{d}_{\rho})}{\text{Tr}(\rho^{\alpha})}- \frac{\text{Tr}
(\rho^{\alpha-\frac{1}{2}}C_{\rho}^{c\dagger})}{\text{Tr}(\rho^{\alpha})} \text{Tr}(\sqrt{\rho}C_{\rho}^{d})-\frac{\text{Tr}(\rho^{\alpha-\frac{1}{2}}C_{\rho}^{d})}{\text{Tr}(\rho^{\alpha})} 
\text{Tr}(\sqrt{\rho}C_{\rho}^{c\dagger})+\text{Tr}(\sqrt{\rho}C^{c^{\dagger} 
}_{\rho})\text{Tr}(\sqrt{\rho}C^{d}_{\rho})=\tilde{\gamma'}^{cd}_{\rho}(\alpha)+\mathbf{i}
\tilde{\sigma'}^{cd}
_{\rho}(\alpha)$. Apart from these two generalizations, one can in principle, deduce other generalizations of the Fubini-Study metric for mixed states. All of these generalizations reduce to Eq. (\ref{22}) for pure states. The Cramer-Rao bound must be naturally satisfied by these generalizations. Depending on the definitions of the operator $C^{i}_{\rho}$, we can again define various Fubini-Study metric tensors on the mixed states with the same properties as before. We can again impose the monotonicity condition and unitary invariance \cite{petz} on this metric to get the Fisher information metric. Generalizing the cencov's uniqueness theorem of classical Fisher information, in the quantum case, it has already been shown by Petz \cite{petz} that all monotone metric can be expressed by the operator concave functions. In the last section, we characterized the monotone square root super-operators and related it with the logarithmic super-operators \cite{petz,luo1}. We showed that square-root super-operators can be expressed by either operator concave or operator convex functions unlike logarithmic derivative super-operators. Below we study the properties of the generalized Fisher information by imposing monotonicity condition on one set of these generalized FS metrics, $\tilde{G}^{cd}_{\rho}(\alpha)$.

\section{Properties of generalized metrics}Here, we list a number of differences between the FS metric for mixed states or the square-root  Fisher information derived from the FS metric in the previous sections and their generalizations in this section. FS metric for mixed states and the square-root Fisher information metric \cite{luo1} reduce to the FS metric for pure states \cite{pati,karol,anandan} when pure states are considered and they both reduces to the classical Fisher information metric in commutative case. But their generalizations although reduce to the FS metric for pure states when pure states are considered but they do not reduce to the classical Fisher information for $\alpha>1$. Instead, numerator of the second metric, $\tilde{G}^{cd}_{\rho}(\alpha)$ reduces to the generalized classical (2,$\lambda$)$^{th}$ Fisher information \cite{lutwak1,lutwak2} in the commutative case, whereas, the first one, $G^{cd}_{\rho}(\alpha)$ has no classical counter part. From here it is easy to get the quantum generalization of classical (p,$\lambda$)$^{th}$ Fisher information \cite{lutwak1,lutwak2} by introducing the first derivative dependence on the relation between the two bases $\{A_{i}\}$ and $\{B_{i}\}$. It has been shown in the previous sections that the super-operators corresponding to the square-root derivatives can be expressed in terms of operator concave or convex functions $f(x)$. We know that the unital CPTP map satisfies the Cauchy Schwartz inequality, i.e, ${\cal E}(A^{*}A)\geq|{\cal E}(A)|^2$ for all normal matrices $A$. Using the same techniques as before and the Cauchy-Schwartz inequality, it can be easily shown that the same characterization holds for $\tilde{G}^{cd}_{\rho}(\alpha)$ here, whereas it is still unclear for $G^{cd}_{\rho}(\alpha)$. One difference with the square-root Fisher information as in \cite{luo1} is that on both the metric spaces $G^{cd}_{\rho}(\alpha)$ and $\tilde{G}^{cd}_{\rho}(\alpha)$ the dynamical phase \cite{pati}, i.e., $i\langle\psi_{AB}|d\tilde{\psi}_{AB}\rangle=i\text{Tr}(\rho^{\alpha-\frac{1}{2}}C^{i}_{\rho})$ is non-zero, because
\begin{eqnarray}\label{28}
\text{Tr}(\rho^{\alpha-\frac{1}{2}}(\mathbf{\theta})C_{\rho}^{i})&=&\sum_{i,j}\rho^{\alpha-\frac{1}{2}}(\mathbf
{\theta})_{ij}[C_{\rho}]_{ji}\nonumber\\&=&\sum_{i}\lambda_{i}^{\alpha-\frac{1}{2}}\frac{d\rho_{ii}}{\sqrt{\lambda_{i}}}\neq 0.
\end{eqnarray}
Although, the dynamical phase is non-zero, the second generalization, $\tilde{G}^{cd}_{\rho}(\alpha)$ is independent of the dynamical phase. An important point to mention here is that both of these generalizations of the FS metric for mixed states must satisfy the Cramer-Rao bound \cite{cramer,fisher,amari,amari1,petz,fuji,matsumoto} as they were derived from the FS metric for pure states.

\section{Conclusion}
%%%%%%%%%%%%%%%%%%%%%%%%%%%%%%%%%%%%%%%%%%%%%%%%%%%%%%%%%%%%%%%%%%%%%%%%%%%%%
In this paper, we have derived the FS metric for mixed states. We connected this metric with the quantum uncertainty as defined in \cite{luo}. Like the FS metric for pure states, it also captures only the quantum part of the uncertainty in the evolution Hamiltonian. This is the most natural generalization of the Fubini-Study distance for mixed states in the sense that it is a $U(1)$ gauge invariant distance along any parameter θ for mixed states unlike the FS metric for pure states, in which case the distance is along the unitary orbit only. This metric satisfies the Cramer-Rao bound. Importance of our metric is that unlike Fisher information metric, it can be measured in the interference experiment. As a result, this novel quantity must be more useful in quantum precision measurement and quantum metrology than the quantum Fisher information. By demanding the monotonicity of the metric, we find the monotonicity condition for the square-root derivative super-operators. We have shown that not only operator concave functions but also operator convex functions give rise to the square-root derivative Fisher information metric. We have shown that along the geodesic path of the Fisher information metric, the dynamical phase \cite{pati} is zero. We generalized the Fubini-Study metric for mixed states further by using the fact that when a mixed state $\rho$ evolves to a final state, purification of the initial state in the extended Hilbert space may evolve to any of the purifications of the final state. We call this metric as $\alpha$ metric. The Cramer-Rao inequality is naturally satisfied by this metric. By imposing the monotonicity and the unitary invariance condition on the super-operator of $\alpha$ metric, we get a new kind of monotone quantum Fisher information metric $G_{\rho}^{ij}(\alpha)$ and $\tilde{G}_{\rho}^{ij}(\alpha)$ which we call as monotone $\alpha$ information metric. We also showed that along a geodesic path of such monotone metric, unlike the previous case the dynamical phase \cite{pati} is non-zero.
%%%%%%%%%%%%%%%%%%%%%%%%%%%%%%%%%%%%%%%%%%%%%%%%%%%%%%%%%%%%%%%%%%%%%%%%%%%%%%
%%%%%%%%%%%%%%%%%%%%%%%%%%%%%%%%%%%%%%%%%%%%%%%%%%%%%%%%%%%%%%%%%%%%%%%%%%%%%%
%%%%%%%%%%%%%%%%%%%%%%%%%%%%%%%%%%%%%%%%%%%%%%%%%%%%%%%%%%%%%%%%%%%%%%%%%%%%%%

%%%%%%%%%%%%%%%%%%%%%%%%%%%%%%%%%%%%%%%%%%%%%%%%%%%
%%%%%%%%%%%%%%%%%%%%%%%%%%%%%%%%%%%%%%%%%%%%%%%%%
%{\em Note added.---}We recently came across the refs. \citep{fungone,fungtwo,taddei,pkok,marcin} which are also devoted to same direction but our %formalism is completely different. An improved and tighter Chau bound also was introduced.

\bibliographystyle{h-physrev4}
%\bibliography{uli}

\begin{thebibliography}{70}
\bibitem{petz}
D. Petz,
\newblock Monotone metrics on matrix spaces, Linear Algebra Appl. {\bf 244} (1996 ) 81-96.
\bibitem{fuji}A. Fujiwara, H. Nagaoka, PLA {\bf 201} (1995) 119-124.\bibitem{matsumoto}K Matsumoto, (2002) J. Phys. A: Math. Gen. 35 3111.\bibitem{amari}S. Amari, Differential-Geometrical Methods in Statistics,
Springer, New York (1985).

\bibitem{amari1}S. Amari and H. Nagaoka, Methods of Information Geometry (Ame. Math. Soc. and Oxford Univ. Press, 2000).
\bibitem{fisher}R. A. Fisher, Theory of statistical estimation, Proceedings of the Cambridge Philosophical Society 22, 700 (1925).
\bibitem{cramer}H. Cramer, Mathematical Methods of Statistics (Princeton University, Princeton, NJ, 1946).
\bibitem{nielsen}M. Nielsen and I. Chuang, Quantum Computation and Quantum Information, Cambridge University Press, (2000).
\bibitem{karol}I. Bengtsson and K. Zyczkowski, Geometry Of Quantum
States, Cambridge University Press, (2006).
\bibitem{anandan}J. Anandan and Y. Aharonov, Phys. Rev. Lett. 65, 1697 (1990).
\bibitem{pati}A. K. Pati, Phys. Lett. A {\bf 159}, 105 (1991). 
\bibitem{luo1}S. Luo, (2007) Commun. Theor. Phys. {\bf 47} 597.
\bibitem{hels}C.W. Helstrom, Phys. Lett. A {\bf 25} (1967) 101.
\bibitem{hels1}C.W. Helstrom, Quantum Detection and Estimation The-
ory, Academic Press, New York (1976).
\bibitem{holevo}A.S. Holevo, Probabilistic and Statistical Aspects of Quan-
tum Theory, North Holland, Amsterdam (1982).
\bibitem{yuen}H. Yuen and M. Lax, IEEE Trans. Inform. Theo. {\bf 19}
(1973) 740.
\bibitem{barn}O.E. Barndorff-Nielsen, R.D. Gill, and P.E. Jupp, J. R.
Stat. Soc. Ser. B {\bf 65} (2003) 775.
\bibitem{luo}S. Luo, Theor. Math. Phys. {\bf 143}, 681 (2005).
\bibitem{cheng}R. Cheng, arXiv:1012.1337.
\bibitem{watanabe}Y. Watanabe, Monotone Metrics on Matrix Spaces, Apr. (2011). This reference is presently unavailable online.

\bibitem{deba} D. Mondal, A. K. Pati, arxiv:1403.5182v2.
\bibitem{supp}see appendix.
\bibitem{wigner}E. P. Wigner and M. M. Yanase, Proc. Nat. Acad. Sci. USA, {\bf 49}, 910 (1963).
\bibitem{lutwak2}E. Lutwak, D. Yang and G. Zhang, IEEE Trans. Inf. Theory, Vol.{\bf 51}, No. 2, Feb. (2012).

\bibitem{tamm} L. Mandelstam and I. G. Tamm, J. Phys. (Moscow) 9, 249
(1945).

\bibitem{Erco}E. Ercolessi, M. Schiavina, arxiv:1205.2561v3.
\bibitem{ecg}P. Facchi, R. Kulkarni, V.I. Man’ko, G. Marmo, E.C.G. Sudarshan, F. Ventriglia, Phys. Lett. A 374 (2010) 4801–4803.
\bibitem{lutwak1}E. Lutwak, S. Lv, D. Yang and G. Zhang, IEEE Trans. Inf. Theory, Vol.{\bf 58}, No. 3, Mar. (2012).


\end{thebibliography}

%%%%%%%%%%%%%%%%%%%%%%%%%%%%%%%%%%%%%%%%%%%%%%%%%%%%%%%%%%%%%%%%%%%%%%%%%%%%%%
$\newline$
$\newline$

{\hspace{3cm}\large \bf Appendix-A}
$\newline$

{\em \bf Operator $\sigma$-means.---}
%%%%%%%%%%%%%%%%%%%%%%%%%%%%%%%%%%%%%%%%%%%%%%%%%%%%%%%%%%%%%%%%%%%%%%%%%%%%%%
Here we define the operator $\sigma$-means as opposite to what is known as operator means \cite{petz,watanabe}. We define operator $\sigma$-means as a binary operations on positive operators such that 
\begin{eqnarray}\label{aa}
&&\sigma:{\cal L}(\cal H)\oplus{\cal L}(\cal H)\rightarrow {\cal L}(\cal H)\nonumber\\&&(A,B)\rightarrow A\sigma B
\end{eqnarray}
if the following conditions are satisfied:

(i)Monotonicity: If $A\leq A'$ and $B\leq B'$ then
\begin{eqnarray}\label{ab}
A\sigma B\geq A'\sigma B'.
\end{eqnarray}

(ii) Transformer Inequality: 
\begin{eqnarray}\label{ac}
C^{\dagger}(A\sigma B)C\geq (C^{\dagger}AC)\sigma (C^{\dagger}BC).
\end{eqnarray}
The transformer inequality implies that

(ii') If C is invertible,
\begin{eqnarray}\label{ad}
C^{\dagger}(A\sigma B)C= (C^{\dagger}AC)\sigma (C^{\dagger}BC).
\end{eqnarray}
The following theorem relates the operator $\sigma$-means and the operator convex functions.

{\bf Theorem.---} For each operator $\sigma$-means $\sigma$, and each operator convex functions $f\geq 0$ on $[0,\infty)$, there are one-to-one correspondence such as $A\sigma B=A^{1/2}f(A^{-1/2}BA^{-1/2})A^{1/2}$.

$Proof.${---} To show the one-to-one correspondence, I followed the lecture note \cite{watanabe}. We need to show that any operator $\sigma$-mean $\sigma$ can be expressed by an operator convex function $f$ as well as any operator convex function $f$ can be expressed in terms of a binary operation known as operator $\sigma$-mean $\sigma$. First we show that any operator $\sigma$-mean $\sigma$ can be expressed by an operator convex function $f$. Let A and B commute with a projector P and since $PAP=AP\leq A$,
\begin{eqnarray}\label{ae}
P(A\sigma B)P&\geq &(PAP)\sigma (PBP)\nonumber\\&=&(AP)\sigma (BP)\nonumber\\&\geq &A\sigma B.
\end{eqnarray}
Then $P(A\sigma B)P-A\sigma B$ is positive and its non-diagonal block is zero, hence
\begin{eqnarray}\label{af}
&&(I-P)(P(A\sigma B)P-A\sigma B)P=0\nonumber\\&\implies &[A\sigma B,P]=0.
\end{eqnarray}
And $[(AP)\sigma(BP),P]=0$.

Multiplying P to Eq.(\ref{ae}), we obtain 
\begin{equation}\label{ag}
(AP)\sigma(BP)P=(A\sigma B)P.
\end{equation} 
So, we can define a new scalar function as $f(t):=I\sigma t$ for $t>0$. Let $A\leq B$, and $A=\sum_{i}a_{i}P_{i}$ and $B=\sum_{i}b_{i}Q_{i}$ such that $a_{i}$, $b_{i}> 0$ for all i and $P_{i}$ and $Q_{i}$ are projectors such that $\sum_{i}P_{i}=\sum_{i}Q_{i}=I$. Then from Eq. (\ref{ag}),
\begin{eqnarray}
I\sigma A&=&(I\sigma A)\sum_{i}P_{i}\nonumber\\&=&\sum_{i}\{P_{i}\sigma(a_{i}P_{i})\}P_{i}\nonumber\\
&=&\sum_{i}(I\sigma a_{i})P_{i}\nonumber\\&=&\sum_{i}f(a_{i})P_{i}\nonumber\\&=&f(A)
\end{eqnarray}
and $I\sigma B=f(B)$. So, the monotonicity condition of the operator $\sigma$-means implies $f(A)\geq f(B)$ and thus $f$ is  an operator convex function. From the transformer inequality we obtain,
\begin{eqnarray}
A\sigma B&=&(A^{1/2}IA^{1/2})\sigma(A^{1/2}A^{-1/2}BA^{-1/2}A^{1/2})\nonumber\\
&=&A^{1/2}\{I\sigma (A^{-1/2}BA^{-1/2})\}A^{1/2}\nonumber\\&=&A^{1/2}f(A^{-1/2}BA^{-1/2})A^{1/2}.
\end{eqnarray}

Now we show that any operator convex function $f$ can be expressed by a binary operation known as operator $\sigma$-mean $\sigma$. Let $A\leq A'$ and $B\leq B'$. We define $X=A^{-1/2}BA^{-1/2}, Y=A'^{-1/2}BA'^{-1/2}, K=A^{1/2}A'^{-1/2}$.
Since $||K||\leq1$, then from the properties of operator convex function we get
\begin{eqnarray}
f(Y)=f(K^{\dagger}XK)\leq K^{\dagger}f(X)K.
\end{eqnarray}
Therefore, we obtain the following monotonicity:
\begin{eqnarray}
A\sigma B&=&A^{1/2}f(A^{-1/2}BA^{-1/2})A^{1/2}\nonumber\\&\geq & A'^{1/2}f(A'^{-1/2}BA'^{-1/2})A'^{1/2}\nonumber\\&\geq & A'^{1/2}f(A'^{-1/2}B'A'^{-1/2})A'^{1/2}\nonumber\\&=&A'\sigma B'.
\end{eqnarray}
Since any operator $K$ such that $A=K^{\dagger}K$ can be written in the form of a polar decomposition $K=UA^{1/2}$, where $U$ is unitary, then
\begin{eqnarray}
K^{\dagger}f[(K^{\dagger})^{-1}BK^{-1}]K&=&A^{1/2}f(A^{-1/2}BA^{-1/2})A^{1/2}
\nonumber\\&=&A\sigma B.
\end{eqnarray}
So, for any invertible C, if $K=A^{1/2}C$, then 
\begin{eqnarray}\label{ah}
(C^{\dagger}AC)\sigma(C^{\dagger}BC)&=&K^{\dagger}f[(K^{\dagger})^{-1}C^{\dagger}
BCK^{-1}]K\nonumber\\&=&C^{\dagger}A^{1/2}f(A^{-1/2}BA^{-1/2})A^{1/2}C
\nonumber\\&=&C^{\dagger}(A\sigma B)C.
\end{eqnarray}
For any projector P, define $K=A^{1/2}P$, and $K'=(PAP)^{-1}PA^{1/2}$, where $(PAP)^{-1}$ is the inverse of PAP for its support. These satisfy
\begin{eqnarray}
&&K^{\dagger}K=PAP,\\&&K'K'^{\dagger}=(PAP)^{-1},\\&&K'K=P,\\&&||KK'||\leq1.
\end{eqnarray}
So, We get
\begin{eqnarray}\label{ai}
(PAP)\sigma(PBP)&=&K^{\dagger}f[K'^{\dagger}PBPK']K\nonumber\\&=&K^{\dagger}
f[K'^{\dagger}K^{\dagger}A^{-1/2}BA^{-1/2}KK']K\nonumber\\&\leq & K^{\dagger}K'^{\dagger}K^{\dagger}f(A^{-1/2}BA^{-1/2})KK'K\nonumber\\&=&
PA^{1/2}f(A^{-1/2}BA^{-1/2})A^{1/2}P\nonumber\\&=&P(A\sigma B)P.
\end{eqnarray}
Using Eq. (\ref{ah}),(\ref{ai}), we can easily show the transformer inequality for operator $\sigma$-means $\sigma$. So, any operator convex function $f$ can be expressed by a binary operation known as operator $\sigma$-mean $\sigma$.
%%%%%%%%%%%%%%%%%%%%%%%%%%%%%%%%%%%%%%%%%%%%%%%%%%%%%%%%%%%%%%%%%%%%%%%%%%%%%%%
$\newline$
$\newline$

{\hspace{3cm}\large \bf Appendix-B}
$\newline$

%%%%%%%%%%%%%%%%%%%%%%%%%%%%%%%%%%%%%%%%%%%%%%%%%%%%%%%%%%%%%%%%%%%%%%%%%%%%%%%%%%
Let us consider two purifications \cite{nielsen,karol} $|\Psi_{AB}\rangle=(\sqrt{\rho(\boldsymbol{\theta})}V_{A}\otimes 
V_{B})\vert\alpha\rangle$ and $\tilde{|\Psi}_{AB}\rangle=(\sqrt{\rho(\boldsymbol{\theta})}V_{A}\otimes 
V_{B})\vert\beta\rangle$ of the same state $\rho$ in the extended Hilbert space, where $|\alpha\rangle=\sum_{i}|A_{i}A_{i}\rangle$ and $|\beta\rangle=\sum_{i}|B_{i}B_{i}\rangle$. Here, we derive the non trivial conditions on the second basis for which both the purifications are the same. We know that
\begin{eqnarray}\label{a1}
&&\langle\Psi|\tilde{\Psi}\rangle =\sum_{i,j}\langle A_{i}A_{i}|
(V_{A}^{\dagger}\sqrt{\rho}\otimes V_{B}^{\dagger})(\sqrt{\rho}V_{A}\otimes V_{B})|B_{j}B_{j}\rangle\nonumber\\&=&\sum_{i,j}\langle A_{i}|V_{A}^{\dagger}\rho V_{A}|B_{j}\rangle\langle A_{i}|B_{j}\rangle=\sum_{i,j}\langle A_{i}|\tilde{\rho} |B_{j}\rangle\langle A_{i}|B_{j}\rangle\nonumber\\&=&\sum_{i,j,k}\langle A_{i}|B_{k}\rangle\langle B_{k}|\tilde{\rho}|B_{j}\rangle\langle A_{i}|B_{j}\rangle,
\end{eqnarray}
such that $\tilde{\rho}=V_{A}^{\dagger}\rho V_{A}$. Now we need to find out under what condition on the basis $\{|a\rangle\}$, $\langle\tilde{\Psi}|\Psi\rangle=1$. We can easily show that 
\begin{eqnarray}\label{a2}
\text{Tr}(\tilde{\rho}^2)=\sum_{i,j}\tilde{\rho}_{ij}\tilde{\rho}_{ji}.
\end{eqnarray}
Comparing Eq. (\ref{a1}),(\ref{a2}), we get $\langle A_{i}|B_{k}\rangle\langle A_{i}|B_{j}\rangle=\frac{\tilde{\rho}_{jk}}{\text{Tr}(\rho^2)}$. There is no unique condition on the basis state $\{|B\rangle\}$. We call this condition on as $\Lambda_{\tilde{\rho_{jk}}}^{2}$. Similarly, other solutions can also be found using the following trace equality as
\begin{eqnarray}
\text{Tr}(\tilde{\rho}^\alpha)&=&\sum_{i_{1},i_{2}...i_{\alpha}}\tilde{\rho}_{i_{1}i_{2}}
\tilde{\rho}_{i_{2}i_{3}}...\tilde{\rho}_{i_{\alpha}i_{1}}.
\end{eqnarray}
Therefore, the other solutions are given by $\langle A_{i}|B_{k}\rangle\langle A_{i}|B_{j}\rangle=\frac{\tilde{\rho}_{ji_{3}}...\tilde{\rho}_{i_{\alpha}k}}{\text{Tr}(\rho^\alpha)}$, where $\alpha=2,3,...n$ and we call them $\Lambda^{\alpha}_{\tilde{\rho_{jk}}}$. For $\alpha=1$, $\Lambda^{\alpha=1}_{\tilde{\rho_{jk}}}=\delta_{ij}$. 

Here, one could also introduce the first derivative dependence in the relation between the bases $\{A_{i}\}$ and $\{B_{i}\}$.
%\section{ Acknowledgement} I thank my advisor Prof. Arun K. Pati for spending his valuable time discussing various issues during the preparation of this manuscript. I also thank him for giving me enough scope and opportunity of independent research.
\end{document}